\documentclass[]{aa}
\usepackage{natbib}
\usepackage{graphicx}
\usepackage{wasysym}
\usepackage{txfonts}
\usepackage{color}
\def\fH2{\mbox{f$_\HH$}}

\def\AF{\mbox{A$_{F\prime,F}$}}

\def\AV{\mbox{A$_{\rm V}$}}

\def\nH2{\mbox{${\rm n}_\HH}$}

\def\pccc{~{\rm cm}^{-3}} 
 
\def\pcc {~{\rm cm}^{-2}}

\def\Tsub#1 {\mbox{${\rm T}_{\rm #1}$}}
\def\TK  {\Tsub K }

\def\arcsec{\mbox{$^{\prime\prime}$}} \def\arcmin{\mbox{$^{\prime}$}}

\def\degr{$^{\rm o}$}
\def\p{\mbox{$^+$}}

\def\cch{\mbox{C$_2$H}}

\def\h13cop{\mbox{{H$^{13}$CO\p}}}

\def\C3H{\mbox{C$_3$H}}
\def\c3h2{\mbox{C$_3$H$_2$}}
\def\cc3h2{\mbox{{\it c}-C$_3$H$_2$}}
 
 \def\R0{R$_0$}
\def\G0{\mbox{G$_0$}} 
  \def\kpc{\rm kpc}
  
\def\ddeg{{}^\circ\kern-.1em}

\def\kms{\mbox{km\,s$^{-1}$}}
\def\ps{\mbox{s$^{-1}$}}
\def\bll{BL Lac}

\def\E#1 {$10^{#1}$}
\def\E#1 {E{#1}}
\def\P#1,{$\nH2\TK~=~#1\times~10^4\pccc$~K}
\def\ec#1,#2,#3,{#1\,(#2)\E{#3}}

\def\H3{\mbox{H$_3$}}

\def\RH2{\mbox{R$_{\rm G}$}}
\def\g13{\mbox{g$_{13}$}} 

\def\cc3h{\mbox{{\it c}-\C3H}}
\def\lc3h{\mbox{{\it l}-\C3H}}

\def\hthcop{\mbox{H$^{13}$CO\p}}

\newcommand{\emm}[1]{\ensuremath{#1}}   
\newcommand{\emr}[1]{\emm{\mathrm{#1}}} 

\newcommand{\hcop}{\emr{HCO^+}} 
 
\newcommand{\HH}{\emr{H_2}}

\renewcommand{\coth}{\emr{^{13}CO}}

\def\CFP{CF\p }

\title{Widespread Galactic CF$^+$ absorption: \\
      detection toward W49 with the Plateau de Bure Interferometer}

\author{H. S. Liszt\inst{1}, V. V. Guzm{\'a}n\inst{2}, J. Pety\inst{3,4}, M. Gerin\inst{4,3}, 
D. A. Neufeld\inst{5} 
 and P. Gratier\inst{6,7}}
\institute{Rational Radio Astronomy Observatory,
           520 Edgemont Road,
           Charlottesville, VA,
           USA 22903
\and       Harvard-Smithsonian Center for Astrophysics, 
           60 Garden Street, 
           Cambridge, MA 02138, 
           USA
\and       Observatoire de Paris (CNRS UMR 8112), 
           61 av. de l'Observatoire, 75014, Paris, 
           France
\and       LERMA/LRA, Ecole Normale Sup\'erieure, 
           24 rue Lhomond, 75005 Paris, 
           France
\and       Johns Hopkins University, 
           Baltimore, MD
           USA 21218
\and       Univ. Bordeaux, LAB, UMR 5804, 
           F-33270, Floriac, 
           France
\and       CNRS, 
           LAB, UMR 5804, 
           F-33270, Floriac, 
           France
}

\begin{document}
\date{received \today}%
\offprints{H. S. Liszt}%
\mail{hliszt@nrao.edu}%
%
\abstract
{}
 {To study the usefulness of \CFP\ as a tracer of the regions where C\p\ and \HH\ 
 coexist in the interstellar medium.}
{We used the Plateau de Bure Interferometer to synthesize \CFP\ J=1-0 absorption
at 102.6 GHz toward the core of the distant HII region W49N at l = 43.2\degr, 
 b=0.0\degr, and we modeled the fluorine chemistry in diffuse/translucent molecular gas.}  
 {We detected \CFP\ absorption over a broad range of velocity showing that \CFP\ is
 widespread in the \HH-bearing Galactic disk gas.}
 {Originally detected in dense gas in the Orion Bar and Horsehead PDR, \CFP\ was 
  subsequently detected in absorption from diffuse and translucent clouds
 seen toward \bll\ and 3C111.  Here we showed that \CFP\ is distributed 
 throughout the diffuse and translucent molecular disk gas with N(\CFP)/N(\HH) 
 $= 1.5-2.0\times10^{-10}$, increasing to N(\CFP)/N(\HH) $= 3.5\times10^{-10}$
 in one cloud at 39 \kms\ having higher N(\HH) $\approx 3\times10^{21}\pcc$. 
  Models of the fluorine chemistry reproduce the observed column densities and 
 relative abundance of HF, from which \CFP\ forms, but generally overpredict the 
 the column density of \CFP\ by factors of 1.4-4. We show that a free space 
 photodissociation rate $\Gamma \ga 10^{-9}\ps$, comparable to that of CH, might 
 account for much of the discrepancy but a recent calculation finds a
 value about ten times smaller.  In the heavily blended and kinematically 
 complex spectra seen toward W49, \CFP\ absorption primarily traces the peaks of 
 the \HH\ distribution.

}


\authorrunning{Liszt, Pety, Guzm{\'a}n, {\it et. al}} \titlerunning{Widespread \CFP\ absorption }

\maketitle{}

%

\section{Introduction}

Recent advances in mm- and sub-mm astronomical spectroscopy open the possibility 
of tracing \HH\ in interstellar gas over a much wider range of column 
density and \HH-fraction, even when carbon monoxide, the usual surrogate tracer 
of \HH\ is not observed.  One new \HH-tracer is HF \citep{SonNeu+10,GerLev+12,SonWol+15}
which is predicted to contain much or most of the gas-phase fluorine even when the
\HH-fraction is small \citep{NeuWol09}.  The abundance of fluorine in
diffuse neutral gas is well-established \citep{SnoDes+07} and the HF/\HH\ ratio 
is relatively insensitive to physical conditions, so that N(HF) is a useful estimator of 
N(\HH).

Ground-state absorption from HF is observed at 1.2 THz.  Observing this 
is a challenge
now, but HF forms the basis for an observable fluorine chemistry 
that is accessible at mm-wavelengths via \CFP\ 
\citep{NeuSch+06,GuzPet+12,GuzRou+12} whose ground state
J=1-0 rotation transition is at 102.6 GHz.   \CFP\ is an immediate descendant 
of HF, formed in the reaction C\p\ + HF $\rightarrow$ \CFP\ + H and destroyed 
by recombination with electrons \citep{NeuWol09}.  Because C\p\ is the dominant 
form of gas-phase carbon well beyond \AV\ = 1 mag \citep{SofLau+04,BurFra+10}, 
and because the \CFP/HF and \CFP/\HH\ ratios are easily modelled, 
\CFP\ may also be a useful tracer of molecular gas where C\p\ is abundant.
It thus has the potential to identify the fraction of C\p\ arising in 
molecular gas.

\CFP\ was discovered in a targeted observation of the Orion Bar 
\citep{NeuSch+06}, subsequently detected in a $\lambda$3mm spectral sweep 
of the Horsehead PDR \citep{GuzPet+12} and then detected in a $\lambda$3mm 
spectral sweep of the absorption spectra of diffuse and translucent clouds 
occulting \bll\ and 3C111 \citep{LisPet+14}. \cite{FecBon+15} recently detected
\CFP\ emission toward a high-mass protostar in a similar spectral sweep and 
used it to infer a source of carbon ionization in the collapsing protostellar 
envelope.

Here we show that \CFP\ is observable in gas seen
along the line of sight to the giant HII region W49N at $l=43.16$\degr, 
$b = 0.02$\degr\ at a distance
of 11.1 kpc \citep{ZhaRei13} and more generally in the gas 
that gathers near the terminal velocity at v $\ga 60$ \kms\ due to 
velocity-crowding \citep{Bur71}.  As a result, \CFP\ is clearly established as a 
commonly-distributed species in diffuse molecular gas.

The structure of this work is as follows.  In Sect. 2 we describe the new and
existing observational material that is discussed.  In Sect. 3 we present the detection
spectrum of \CFP\ toward W49N and compare its kinematics and column densities with those 
of other species seen in absorption along the same sightline.  In Sect. 4 we
compare the column densities of HF and \CFP\ with models  of the fluorine chemistry
in diffuse/translucent clouds and Sect. 5 is a summary.

\begin{figure}
\includegraphics[height=15.1cm]{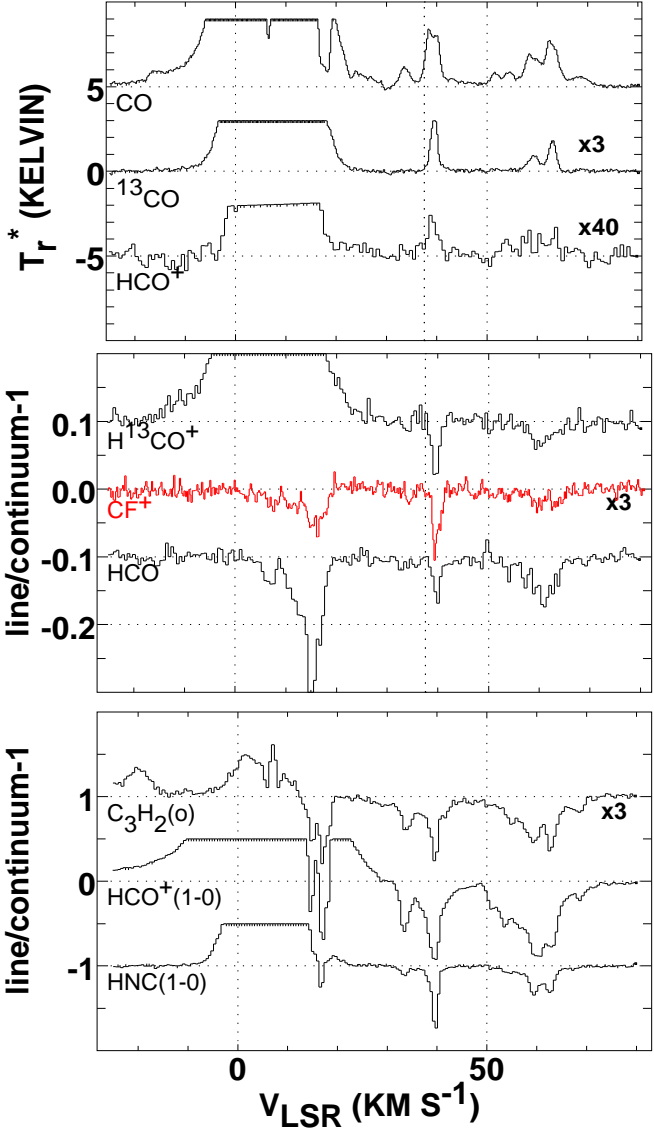}
  \caption[]{Line profiles observed toward W49N.  Top: emission spectra from 
 the ARO 12m antenna at $\approx 1$\arcmin\ spatial resolution.  
 Brighter emission below 20 \kms\ from the W49N H II  region-molecular
 cloud complex has been truncated to allow better viewing of weaker emission
  from the Galactic plane clouds. Middle: absorption from weaker-lined transitions 
including \CFP\ studied at the PdBI.  Bottom: absorption in stronger and more 
  ubiquitously distributed  species observed at the 30m.  The absorption data 
  are plotted as line/continuum ratios, displaced
  as needed to fit the plot. In some cases the profiles have been scaled  
  to make weak features more easily visible, with a scale factor as shown 
  at the right hand side of the spectrum.}
\end{figure}

\section{Observations and data reduction}

\subsection{PdBI absorption measurements toward W49N}

We observed the \CFP\ J=1-0 line at 102.587 GHz (Table 1) with the 
Plateau de Bure Interferometer during 34 hours of telescope time, 
corresponding to 9.3 hours of on-source time scaled to a six antenna 
array, after filtering out low-quality visibilities.  The observations 
were carried out in August and September 2013 in the D configuration, 
achieving a spatial resolution of 5.95\arcsec $\times$ 4.72\arcsec\ so 
that 1 Jy/beam is equivalent to 4.14 K.  The typical precipitable water 
vapor amounted to 6 mm with a typical system temperature of 150 K.  The 
bright quasars 3C454.3 and 3C279 were used to calibrate the bandpass. The 
phase and amplitude temporal variations were calibrated by regularly observing 
two nearby quasars (1827$+$062 and 1923$+$210).  The absolute flux scale 
was derived from observations of MWC349.  The calibration and imaging of 
the PdBI data was done with the \texttt{GILDAS}\footnote{See
\texttt{http://www.iram.fr/IRAMFR/GILDAS} for more information about the
GILDAS software~\citep{Pet05}.}\texttt{/CLIC} and \texttt{MAPPING}
software. The data were gridded onto a 1.24\arcsec\-pixel spatial map 
with rms noise 0.0105 Jy/beam in spectral channels of width, spacing 
and resolution 78.1 kHz or 0.228 \kms.

The 3mm continuum peaked at $\alpha$(J2000) = 19$^h$10$^m$13.45$^s$, 
$\delta$(J2000) = 9\degr06\arcmin13.2\arcsec\ or $l=43.2$\degr,
$b=0.0$\degr\ in Galactic coordinates, with a flux of 3.59 Jy/beam.  
The \CFP\ absorption line profile was extracted at a position 
with a cleaner spectral baseline that was displaced one pixel to the 
West and South having a flux of 3.47 Jy/beam, leading to a
channel-channel rms line/continuum ratio of $2.4\,10^{-3}$.   
The \CFP\ spectrum is shown in the middle panel of Fig. 1.  
The \CFP\ absorption is overlain by the emission profile of an
H66$\epsilon$ recombination line from the ionized gas and the 
procedure for extracting the \CFP\ absorption profile in the 
presence of the recombination line is described in Appendix A.  

\subsection{Other absorption profiles toward W49N}

For comparison with other species observed in the 3mm band we used the 
absorption profiles of HNC and \hcop\ taken at the IRAM 30m telescope
by \cite{GodFal+10}, along with similar profiles of \c3h2, \cch\ 
\citep{GerKaz+11} and \hthcop.  These spectra have 39 kHz channel spacing and 
78 kHz spectral resolution equivalent to 0.117 \kms\ spacing at 100 GHz.  
The HCO spectrum shown in Fig. 1 is from 
\cite{LisPet+14}, shown at 0.270 \kms\ channel spacing and resolution.
For CH we used the hyperfine-deconvolved 532 GHz absorption profile 
from \cite{GerdeL+10} at 0.135 \kms\ spacing and resolution, for HF the 
profile of \cite{SonNeu+10} at 0.141 \kms\ spacing and resolution
 and for C\p\
a smoothed absorption spectrum from \cite{GerRua+15} at 0.63 \kms\ spacing 
and resolution.
The sub-mm profiles were taken as as part of the Herschel 
PRISMAS Project.  The $\lambda$3mm profiles are shown in Fig. 1 
and with the spectrum of CH in Fig. 2.  Profiles of the  C\p\ and HF 
absorption lines can be seen in Fig. 4.  Line frequencies
and optical depth-column density conversions for these species
are given in Table 3.

When profiles are compared on a channel-by-channel basis the narrower of the
two was convolved to the spectral resolution of the other using a gaussian
kernel and then regridded using  cubic splines to create a common velocity axis.

\subsection{CO and \hcop\ J=1-0 emission}

\begin{figure}
\includegraphics[height=16cm]{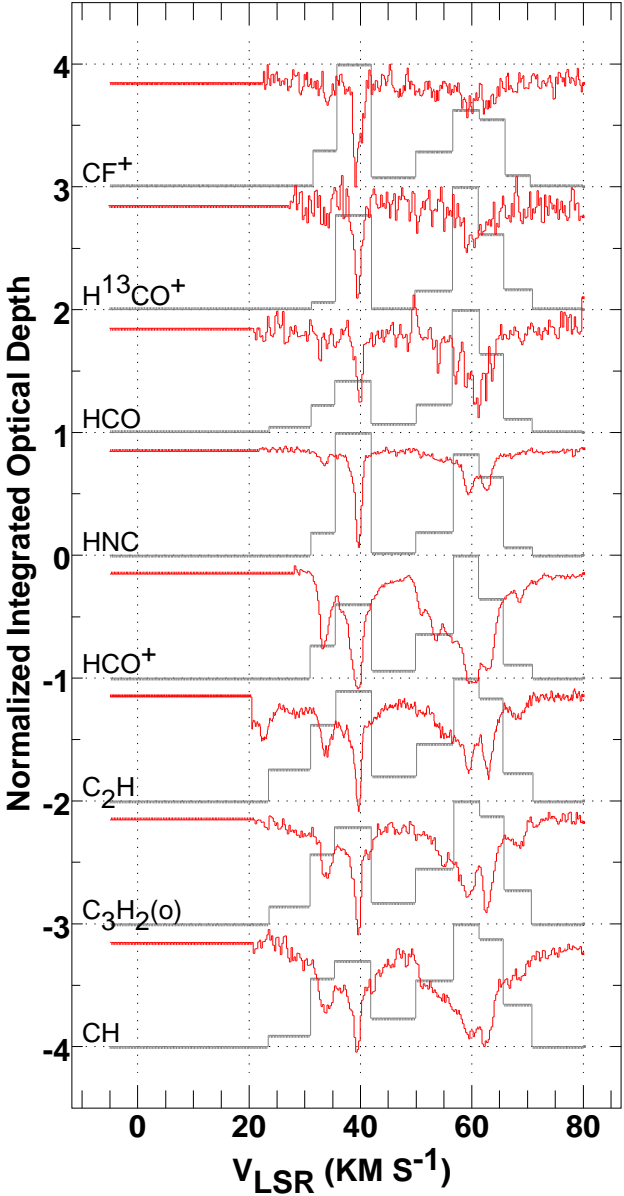}
  \caption[]{ Histograms of integrated optical depth over discrete velocity 
  ranges for species observed in absorption toward W49N, normalized to a 
  maximum of unity in each case. The quantities plotted are those given in 
  Table 2.  Scaled and offset
  absorption spectra are shown superposed in red to illustrate the underlying
 kinematic features in the velocity bins.}
\end{figure}

The emission profiles shown in the top panel of Fig. 1 were acquired
at the original ARO 12m telescope during April of 2011 by position
switching against a distant off position whose spectrum was determined
by position switching against another similarly distant off-position 
and added back to the data.  The beamwidth of this antenna was 1\arcmin\
at 110 GHz, scaling inversely with frequency.
The temperature scale of the ARO 12m telescope, like that of the NRAO
12m telescope before it, was T$_r^*$, approximately 0.85 T$_{\rm mb}$.
The emission spectra have spectral resolution and channel spacing 48.8 kHz.
The \hcop\ spectrum shown in Fig. 1 is the average of four spectra taken
at positions displaced $\pm1.4$\arcmin\ to the East and North, to avoid
contamination of the very weak emission by absorption.  The CO profiles 
were taken directly toward the continuum, as the W49N continuum was a small 
fraction of a Kelvin at the ARO 12m antenna.  Reliably observing carbon 
monoxide in absorption toward W49N would require spatial resolution 
and resolved flux substantially exceeding those  achieved in the PdBI 
synthesis of \CFP\ here, for which the peak flux of 3.6 Jy/beam was equivalent 
to 15 K.  This is not so much larger than the
4 K emission brightness of CO that a pure absorption spectrum could
be  acquired.

\subsection{Optical depths, error estimates, upper limits and column density 
conversions}

The spectroscopic parameters of the observed \CFP\ transitions are given in 
Table 1 and optical depth integrals for \CFP\ and other species discussed here
are given in Table 2.  Quantities in parentheses in Table 2 are 1$\sigma$ rms 
errors in the tabulated quantities determined from the rms 
noise level at zero absorption but accounting for the saturation in the
absorption line profile.  Where the integrated optical depth was 
detected at a level well below $3\sigma$ the integral is shown as zero and
$3\sigma$ upper limits are plotted in the various Figures.  

In the present experiment the hyperfine structure 
of \CFP\ was unresolved and the integrated optical depth is 50\% larger 
than if the stronger component had been observed alone.  
Hence N(\CFP) is derived from the integrated optical depth in 
Table 2 using N(\CFP) = $2/3 \times 2.02 \times 10^{13} \pcc 
\int \tau dv$ with the optical depth integral expressed in \kms.
Conversion factors between integrated optical depth and column density
are given separately in Table 3, assuming excitation in equilibrium with 
the cosmic microwave background. 

\section{The line of sight toward W49N}

\subsection{Line profiles}

$\lambda$3mm line profiles observed toward W49N are shown in Fig. 1.  
Emission lines from dense gas in the W49N complex peak at 0 - 15 \kms\ but 
may have broad wings extending well beyond this range:  \CFP\ and HCO are 
notable exceptions because they do not appear in emission at the PdBI.  
When the emission is not too broad the continuum and the source emission 
profiles are noticeably absorbed at 5-20 \kms.   Much of this absorption is due to 
gas within and on the near-side of the W49N complex itself: The relationship between 
emission and absorption profiles of HCO and \hthcop\ as seen by the PdBI toward W49N 
is shown in Fig. 4 of \cite{LisPet+14}.  Interstellar absorption from gas lying 
closer to the Sun must also be present at v $\la$ 20 \kms\  but is masked by 
the contribution from W49N. As shown in Appendix B, the velocity range v $<$ 20 \kms\
corresponds to the nearest 1.75 kpc along the line of sight.

\begin{figure*}
\includegraphics[height=7cm]{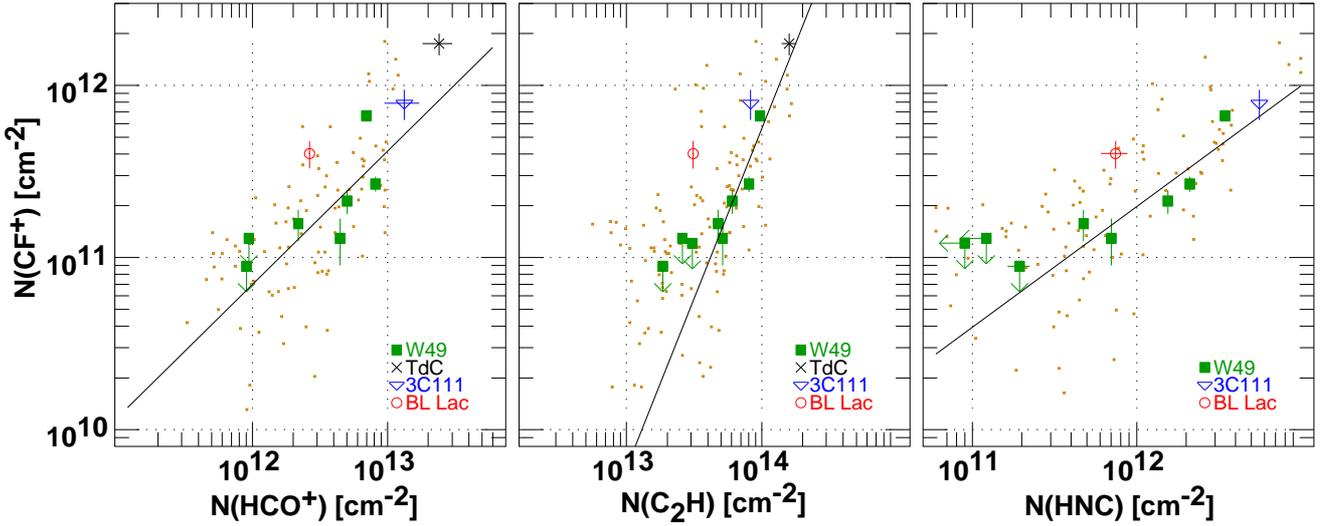}
  \caption[]{Column density of \CFP\ vs. those of \hcop\ (left), \cch\ (middle) 
and HNC (right) for the velocity intervals  in Tables 2 and 4.  Shown for comparison
are data from diffuse cloud sightlines toward \bll\ and 3C111, and for the 
Horsehead (TdC).  The small brown points  are 
 the individual channel values of optical depth in the original spectra, scaled 
 equally in each coordinate to fit in the plot area. Upper limits are 3 $\sigma$.
The solid lines are regression lines fit to the detections toward W49, ignoring the upper 
limits.}
\end{figure*}

At bottom in Fig. 1 are strong interstellar absorption lines of HNC, \hcop\
and \c3h2, representing three distinct chemical families.  \hcop\
\citep{LucLis96} and the hydrocarbons CH \citep{GerdeL+10}, 
\cch\ and \c3h2\ \citep{LucLis00C2H,GerKaz+11}  are especially ubiquitous 
and show the widest range of absorption.  The hydrocarbons and \hcop\
have rather stable abundances with respect to \HH, which we use to
derive column densities of \HH\ using X(CH) = N(CH)/N(\HH) 
$= 3.5\times10^{-8}$ \citep{SheRog+08} toward W49N and X(\hcop) 
$= 3\times10^{-9}$ (Table 4 and \cite{LisPet+10}) for quasar lines of sight 
where CH was not observed.   Scatter
in the CH-\HH\ relationship, uncorrected for measurement errors, is
a factor 1.6 as employed in Fig. 5.  The column
densities of CN, HNC and HCN vary together in fixed proportion but 
only achieve high abundance relative to \HH\ at higher N(\HH)
\citep{LisLuc01,GodFal+10} and presumably at somewhat higher number
density as well.  For this reason they have more limited absorption profiles
but serve to distinguish the more diffuse molecular gas.

In the middle panel of Fig. 1 are profiles of three weaker-lined species
including the detection spectrum of \CFP.  With sufficient sensitivity the 
interstellar \hthcop\ and \hcop\ absorption profiles would presumably be 
identical in shape.  With the present sensitivity all of these species show
absorption over only a limited range, most notably in the 40 \kms\ cloud 
feature, but also in the terminal-velocity ridge above 60 \kms. 

At top in Fig. 1 are J=1-0  emission profiles of CO ($^{12}$CO), \coth\
and \hcop.  It is notable that CO emission is detected at all velocities
at which absorption is observed in the hydrocarbons and \hcop.  
\coth\ is detected over a more limited range, in part due to considerations of 
signal/noise for the weaker line.  The emission at 33 \kms \ has a very high
ratio of CO to \coth\ emission, presumably reflecting optically thin emission and 
a small CO column density.

As shown in the uppermost panel of Fig. 1, \hcop\ emission is detected 
at 39 \kms and v $\ga$ 50 \kms\ at a level $\approx 1-2$\% that of CO 
as is typical of emission observed in the galactic plane in the inner
Galaxy \citep{Lis95,HelBli97}. 
 
\subsection{Line of sight kinematics}

The kinematic characteristics of the absorption profiles toward W49N result
from several influences as illustrated in Fig. B.1 and discussed in Appendix B. 
W49N is 11.1$\pm$0.8 kpc distant from the Sun \citep{ZhaRei13}, hence at 
galactocentric radius R = 7.6 kpc, just inside the Solar Circle on the other 
side of the Galaxy for a Sun-center distance of 8.5 kpc.  Hence lines 
originating within the W49N complex are closest to 0-velocity because of 
its inadvertent proximity to the Solar Circle on the far side of the Galaxy.  
The highest velocities appearing in emission or absorption in Fig. 1 correspond 
quite well to the maximum velocity predicted for a flat rotation curve with 
R$_0$ = 8.5, $\Theta(R)$ = 220 \kms\ as shown in Fig. B.1.  An apparent 
concentration of gas arises at the high-velocity end of the line profiles  
as the result of velocity-crowding \citep{Bur71}, as the slope of the 
velocity-distance relationship approaches 0 at the so-called sub-central 
point where the line of sight passes nearest to the center of the Galaxy.  Gas is
concentrated into the velocity range 30 - 40 \kms, with notable gaps
at higher and lower velocity, owing to sharp changes in the velocity-distance
relationship resulting from  spiral-arm streaming.
This has not been extensively modeled for W49N but a detailed discussion
of the influence of spiral arm streaming along the sightline toward W43
is given in \cite{LisBra+93}. 

\subsection{Column densities and cloud properties}

To give a broad characterization of the properties of the gas lying along the 
sightline to W49N we divided the velocity range of the interstellar absorption 
into a set of intervals based on the apparent features in the 
absorption profiles, and we calculated the integrated absorption optical 
depth for a variety of molecules shown in Fig. 1 and 2, as given in Table 2.
It is tempting to identify narrow-lined spectral features with individual
clouds, especially at 39 \kms, but there is severe blending at all
velocities. 
The species in Table 2 include the three hydrocarbons CH, \cch\ and \c3h2\ 
that are known to track each other \citep{GerKaz+11} and to trace 
\HH, and \hcop\ that  also has a relatively stable relative abundance 
with respect to \HH.  The hydrocarbons and \hcop\
have much more broadly distributed absorption than HNC, for instance, 
even though the peak absorption is not much different.     

The results are shown graphically in Fig. 2 where the histograms of integrated 
optical depth and the line profiles are shown superposed.  In each case the 
histogram is normalized to have a unit peak.  Species which most closely track 
the total column density of \HH\ without regard to the number or column 
densities of individual regions  (the hydrocarbons and \hcop) have higher 
column density over the terminal velocity ridge around v = 60 \kms.  The hydrocarbons 
and \hcop\ also  have a relatively high ratio of absorption at 33 \kms\ compared 
to that at 39 \kms.  HNC, which is known to be more abundant in individual clouds 
of higher \HH\ column density \citep{LisLuc01}, behaves oppositely to the 
hydrocarbons and \hcop.  HCO is an extreme example of a species appearing
preferentially in the more diffuse gas in the terminal velocity ridge.  
\CFP\ is clearly most similar to HNC, which is something of a surprise
given its origin in HF, which tracks \HH, and C\p\ that is ubiquitous
(see Sect. 4 and Fig. 4).


The optical depth integrals in Table 2 are converted to relative abundance 
N(X)/N(\HH) in Table 4 using the column density-optical depth conversion
factors given in Table 3 to form the ratio N(X)/N(CH) and 
N(CH)/N(\HH) $= 3.5\times10^{-8}$.  Shown at the
bottom of Table 4 are two summary quantities:  the mean relative abundance
found by integrating over the profiles of the species listed relative
to the total CH column density, and the mean and rms of the individual
entries in the Table, for those intervals where the species in question
was detected.  As noted by  \cite{GerKaz+11} the hydrocarbons have quite
stable abundances relative to each other:  \cch\ and \c3h2\ have rms
of 20\% or less of their mean table entries.  HF is also relatively stable with
respect to CH, with a fractional rms of 23\%.  The fractional rms of \hcop\ and HCO
are intermediate, 42\%, and those of \CFP\ and HNC are the largest,
48\% and 69\%, respectively.   The overall average abundance X(\hcop) =
$2.7\,10^{-9}$ is consistent with the value X(\hcop) = $3\,10^{-9}$ derived 
from quite  other considerations by \cite{LisPet+10}.  

In all, the results in Table 4 shows an impressive consistency of the diffuse 
cloud chemistry over the Galactic disk, and validates the  empirical methods
that have been used across many spectral domains to extract molecular
column densities and relative abundances.  

\section{The abundance and chemistry of \CFP}

\subsection{N(\CFP) and chemical relationships}

Extending the analysis in \cite{LisPet+14}, Fig. 3 shows the \CFP\ column 
density with respect to \hcop, \cch\ and HNC for W49N and for the sightlines
observed toward compact extragalactic continuum sources \bll\ and 3C111, 
as well as data taken toward the Horsehead PDR. 
Shown for W49N are the integrals tabulated in Table 2 and the cloud of
channel-by-channel datapoints equally scaled in both dimensions so as
to overlay the datapoints while preserving the  slope. 
Shown in each panel is a regression line fit to the W49N detections alone.  
The slopes of these fits are $0.78\pm0.56$ for \hcop,
$1.97\pm0.44$ for \cch\ and $0.70\pm21$ for HNC. 

The velocity ranges chosen toward W49N have column densities of \hcop,
\cch\ and HNC overlapping those seen toward \bll\ and 3C111 and the
regression line fits show a high degree of consistency between  the
W49N sightline, 3C111 and, where possible, the Horsehead PDR.  The
line of sight toward \bll\ is an obvious outlier, with 2-3 times more
\CFP\ compared to the other datapoints.  The 39 \kms\ cloud toward W49N is a
lesser outlier with respect to \hcop.

A striking aspect of Fig 3 is the quadratic slope of the \CFP-\cch\
relationship as compared with the much slower than linear variation 
with respect to HNC.  This is indicative of a rapid variation of 
N(HNC) with N(\cch), as was in fact shown in Fig. 3 of 
\cite{LisLuc01}: N(HNC) increases abruptly by about a factor 100 in
individual cloud features for N(\cch) $\ga 10^{13}\pcc$,
and also for N(\hcop) $\ga 10^{12}\pcc$.  By contrast,
\CFP\ has a faster than linear variation only with \cch.

\begin{figure}
\includegraphics[height=7cm]{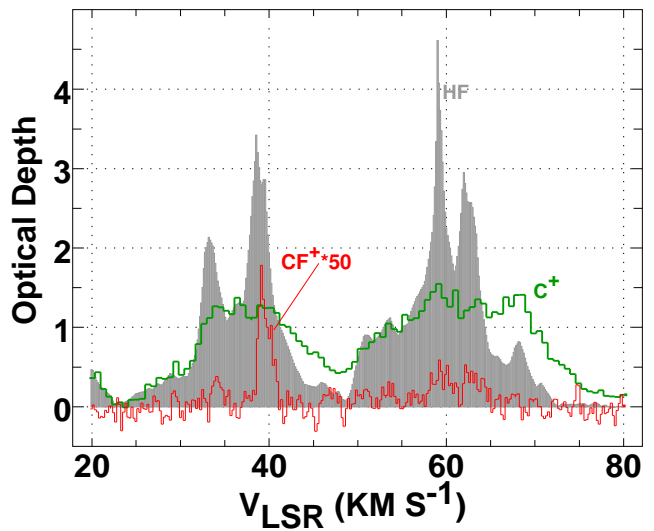}
  \caption[]{Optical depth profiles of \CFP\ (scaled upward by a factor 50)  and the 
species HF and C\p\ that form it.}
\end{figure}

\begin{figure}
\includegraphics[height=13cm]{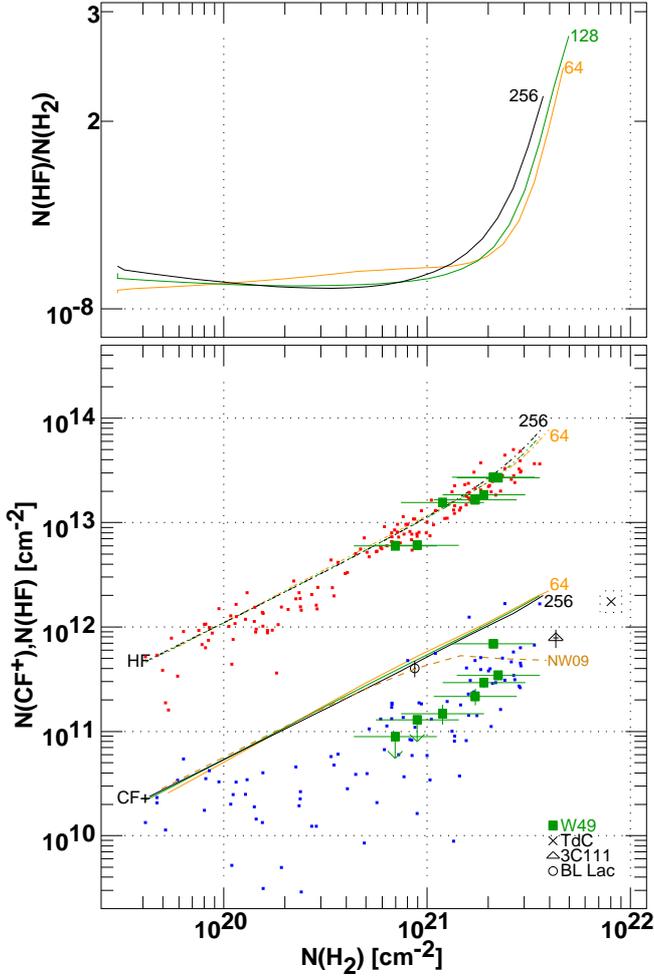}
  \caption[]{Model and observed \CFP\ and HF abundances.  Upper: the HF/\HH\
ratio in models of HF and \HH\ formation at total densities n(H) = 64, 128 and 256 $\pccc$ .
Lower: observed and model HF and \CFP\ column densities.  The W49N data were placed
in the plane using N(CH)/N(\HH) $= 3.5\times 10^{-8}$ and the data for \bll\
and 3C111 were positioned relative to \hcop\ assuming N(\hcop)/N(\HH) = 
$3\times 10^{-9}$.  Profile sums over the velocity intervals in Table 2
are shown as solid symbols and upper limits are 3 $\sigma$. 
Individual channel values of the HF and CF\p\ optical depth spectra
are shown as red and blue  points, respectively, after scaling in each coordinate by the Galactic 
line of sight velocity gradient for a flat rotation curve (see text). 
The dashed light brown curve shows the results of \cite{NeuWol09}.}
\end{figure}

\begin{table}
\caption[]{CF\p\ transitions observed and column density-optical 
depth conversions$^a$}
{
\small
\begin{tabular}{lcccc}
\hline
Frequency &F$^\prime$ - F & J$^\prime$ - J & \AF     & N(X)/$\int\tau dv^b $ \\
 MHz      &        &  & 10$^{-6}$\ps  &   $\pcc (\kms)^{-1}$ \\
\hline
102587.189 & 1/2-1/2 & 1-0 & 4.82 & $4.04\times10^{13} $ \\
102587.533 & 3/2-1/2 & 1-0 & 4.82 & $2.02\times10^{13} $ \\
\hline
\end{tabular}}
\\
$^a$ Spectroscopic data from \cite{GuzRou+12} and {\tt www.splatalogue.net}\\
$^b$ Assuming excitation in equilibrium with the cosmic microwave background \\
\end{table}

%
%

\subsection{Modelling the fluorine chemistry}

The fluorine chemistry of diffuse and translucent gas has been summarized
by \cite{NeuWol09}.  It is dominated by the formation of HF, which is
expected to contain much or most of the free gas-phase fluorine in regions
where the \HH\ fraction is appreciable.  However, as noted by \cite{SonWol+15}, 
the rate constant for the primary reaction governing the formation of HF,
F + \HH\ $\rightarrow$ HF + H, has recently been found to be rather
smaller at low temperature than previously believed \citep{TizLeP+14}.
This complicates discussion of the fluorine chemistry because the 
HF/\HH\ ratio is more variable, and generally smaller, than in the models 
of \cite{NeuWol09}, see \cite{SonWol+15}.  

HF is also formed in a minor way by the reaction of neutral fluorine atoms 
with CH, OH, and, more slowly, with water. In diffuse gas, HF is destroyed 
in reaction with C\p\ (the primary process, forming \CFP), in reaction with He\p\
(created by  cosmic-ray ionization) and by photodissociation.  \CFP\
is destroyed only by electron recombination in the chemical model of
\cite{NeuWol09} that was adopted here.   Because
C\p\ usually dominates both the formation and destruction of \CFP\ in diffuse
gas, the \CFP/HF ratio is generally insensitive to the number density as 
long as C\p\ is the dominant form of gas-phase carbon and provides
the majority of free electrons.

   

The species involved in the \CFP\ formation chemistry are observable and 
Fig. 4 shows the optical depth profiles (equivalently the profiles of
column density or column density per unit velocity) of C\p, HF and \CFP.  
Because it is so widespread and so heavily blended, and to some extent
because it was observed with some three times lower velocity resolution, 
the C\p\ profile varies little across the peaks in HF and \CFP. 
The optical depth of \CFP\ varies more strongly than that HF or
the product $\tau$(C\p)*$\tau$(HF). 

To model the \CFP\ chemistry we adopted a very slightly simplified version of 
the fluorine chemistry of \cite{NeuWol09} using the new HF formation rate of 
\cite{TizLeP+14} and fixed gas-phase elemental abundances 
[C]/[H] = $1.4\times 10^{-4}$ \citep{SofLau+04}, [F]/[H] = $1.6\times 10^{-8}$ 
\citep{SnoDes+07}.  We held fixed the empirically determined molecular abundances
X(CH) = $3.5\times10^{-8}$, X(OH) = $10^{-7}$ \citep{WesGal+10} 
and X(\HH O) = $1.2\times10^{-8}$  \citep{SonWol+15}
and embedded this chemistry in models of the heating and cooling of
uniform density gas spheres, such as were recently used to calculate abundances of 
\HH, HD and CO, etc. \citep{Lis07CO,Lis15HD}: these use  standard values of
the radiation field and photodissocation rates, etc.  We adopted a constant
gas-phase abundance of fluorine as in  \citep{SonWol+15}
and did not include a progressive freeze-out of gas-phase fluorine, which 
apparently occurs only at higher values of the hydrogen column density 
than are encountered in this work, see Fig. 5 and the discussion below.

Model results for HF and \CFP\ are shown in Fig. 5 for three densities 
n(H) = 64, 128 and 256 $\pccc$: the proportions of atomic and molecular
hydrogen are calculated inside the models self-consistently.  The results 
plotted are integrated 
over the central line of sight toward models with varying N(H) following the
H-\HH\ transition across 128 equi-spaced radial shells.  
X(HF) $\approx 1.1\times 10^{-8}$, independent of n(H) for N(\HH) 
$\la 2\times 10^{21}$ or \AV\ $\la 2$ mag.  At higher N(\HH)
carbon recombines to neutral carbon and CO and the
photodissociation rate of HF decreases, allowing 
X(HF) to increase to the point where HF consumes all of the free gas-phase 
fluorine (see also the models of \cite{SonWol+15}).  X(\CFP) remains nearly 
constant as the decrease in C\p\ is compensated by falling electron
density.  The HF/\CFP\ ratio increases, which we have traced in
the models to falling temperature and faster \CFP\ recombination.  

Shown for comparison is the earlier prediction for N(\CFP) of \cite{NeuWol09}, 
which includes progressive freeze-out of the gas-phase fluorine. Agreement of the 
two predictions for \CFP\ is surprisingly good considering the differences in the 
underlying HF chemistry.  The HF column densities in \cite{NeuWol09} are about 
50\% higher than here at N(\HH) $\la 2\times 10^{21}\pcc$.
  
\subsection{Model and observed results for HF}

Shown in Fig. 5 are HF and \CFP\ column densities over  the
velocity intervals tabulated in Table 2 and 4, and  the clouds of channel
by channel optical depths scaled to fit in the plot area. As
noted in Sect. 2, N(\HH) is approximated as N(\HH) = N(CH)/$3.5\times 10^{-8}$,
here shown with an assigned error in N(\HH) of a factor 1.6 corresponding to 
the scatter in the empirical CH-\HH\ correlation of \cite{SheRog+08}.
Also shown in Fig. 5 are our measurements of \CFP, 
along with prior measurements in the Horsehead PDR \citep{GuzPet+12} and in
absorption toward \bll\ and 3C111 \citep{LisPet+14}.  The quasar absorption 
datapoints were placed horizontally by assuming a relative abundance 
N(\hcop)/N(\HH) $= 3\times 10^{-9}$ (see \cite{LisPet+10} and 
Table 4).

The results for HF are well explained by the models, and have a nearly constant 
abundance relative to \HH\ at N(\HH) $\la 2\times10^{21}\pcc$, 
X(HF) = $1.1\times10^{-8}$ in agreement with the results of 
\cite{IndNeu+13} and \cite{SonWol+15}.  As sampled in HF, whose abundance
with respect to \HH\ remains steady up to the highest inferred values
of N(\HH), freeze-out of the fluorine
and/or collapse of the carbon ionization to CO and neutral carbon
must occur at values of N(\HH) beyond  those  encountered along
the line of sight to W49N.  The absorption line of HF becomes heavily saturated 
at higher column densities, see \cite{SonNeu+10}

\subsection{Model and observed results for \CFP}

The predicted value over the range of N(\HH) observed toward W49
is X(\CFP) $= 5.3 \times 10^{-10} \approx$ X(HF)/20.
This can be understood simply by comparing the rate of \CFP\ formation
from C\p + HF $\rightarrow$ \CFP\ + H and the rate of \CFP\ recombination
in a gas at 75 K (the mean temperature of \HH\ \citep{RacSno+02}) 
in which C\p provides 7/8 of the ambient electrons as determined from
detailed examination of the internal structure of the models.

The ratios N(\CFP)/N(\HH) observed in absorption toward W49N (Table 4) are 
$ 3.3\times10^{-10}$ for the gas around 39 \kms\ and $1.40\pm0.16\times10^{-10}$ 
for the other four velocity intervals in which \CFP\ was detected. The smaller
of the observed relative abundances is about a factor four below the model 
prediction while that at 39 \kms\ is about 40\% too small. 
 The line of sight toward \bll\ was already cited as an 
outlier in the discussion of Fig. 3.  The \CFP\ abundance along 
the line of sight through the translucent gas toward 3C111 is consistent with 
the smaller of the  \CFP\ abundances toward W49N.

Agreement of the model and observed results for HF implies that the ratio 
of HF formation and destruction rates is correct under the assumption of 
equilibrium of the chemical rate equations.  Given that the formation 
rate using the recently measured reaction rates of \cite{TizLeP+14} 
should be secure and the H-\HH\ transition is modelled in detail, it 
follows that the destruction rate of HF should also be correct.   
Therefore the formation rate of \CFP\ is also correct to the extent 
that formation of \CFP\ dominates the destruction of HF.  This implies 
that small \CFP\ abundances arise from the absence of some important 
\CFP\ destruction mechanism in the models, and perhaps that this 
mechanism is suppressed in the 39 \kms\ gas toward W49N and toward \bll.

\subsection{A role for \CFP\ photodissociation?}

We considered that this could point to a role for photodissociation,
 which was not incorporated in the underlying chemical model given 
the rapidity with which \CFP\ recombines.  \cite{GuzPet+12} noted that 
even a relatively large free-space photodissociation rate 
$\Gamma = 10^{-9}\ps$ was too slow to affect the abundances seen in 
the Horsehead nebula and they neglected it in 
their calculation of the \CFP\ column density.

To clarify this matter we examined our models to see what were the actual 
rates of photodissociation of \CFP.  The ranges of recombination 
rates in the models averaged along the line of sight were 
$1.8\,10^{-9} .. 0.8\,10^{-9}\ps~(64\pccc), 
3.6\,10^{-9} .. 1.5\,10^{-9}\ps~ (128\pccc)$ and 
$6.9\,10^{-9} .. 1.9\,10^{-9}\ps~ (256\pccc)$
with the higher values at the smallest N(\HH) and vice versa.
Thus a free space photodissociation rate $\Gamma \ga 10^{-9}\ps$ would have
a very noticeable effect on X(\CFP), introducing number and column density 
dependences beyond the very small ones in the models shown here.
This is a relatively high rate, but not higher than that for CH \citep{VanBla86}
which is quoted in the UFDA database \citep{McEWal+13} as $1.7\times10^{-9}\ps $ 
However, we subsequently learned that the free-space photodissociation rate 
of \CFP\ has recently been calculated to be $2\times 10^{-10}$\ps\ 
(Dayou and Roueff, in preparation), effectively eliminating 
photodissociation as a practical solution to the discrepancy between
the calculated and observed \CFP/HF ratio.  

In closing we note that some   uncertainties remain regarding the underlying
chemistry.  The adopted \CFP\ recombination rate was measured in a gas of 
rotationally-warm \CFP, and may need refinement.
Moreover, the assumed formation rate for CF+ by reaction of C+ with HF
has not yet been measured in the laboratory; here, we simply adopted the
capture rate computed by NW09, assuming - in effect - that every capture
results in a reaction.  In reality, the CF+ formation could be
significantly smaller than what we have assumed, providing an alternative
explanation for the lower-than-predicted CF+ abundances reported here.

\section{Summary}

\CFP, formed in the reaction C\p\ + HF $\rightarrow$ \CFP\ + H, was originally
detected in the Orion Bar and subsequently detected in the Horsehead PDR and
then in diffuse/translucent molecular clouds occulting \bll\ and 3C111 and
in a massive protostellar envelope.
In this work we detected the 102.6 GHz J=1-0 lines of \CFP\ (Table 1) 
in absorption from gas lying in the galactic plane toward W49N , a giant HII 
region at a distance of 11.1 kpc from the Sun at $l=43.2$\degr, $b=0.0$\degr.
As detailed in Appendix A the \CFP\ profile was extracted in the presence
of an overlying H66$\epsilon$ recombination line.
As shown in Fig. 1 and 2 and summarized in Tables 2 and 4, narrow absorption 
was seen in features at 33 and 40 \kms\ corresponding to 
well-known discrete diffuse/translucent molecular gas clouds lying in the 
Galactic plane.  But absorption from more diffuse material was also 
observed at intermediate velocities and  in the terminal-velocity ridge 
at v $\ga 60$ \kms\ caused by velocity crowding of unrelated gas parcels 
as the line of sight velocity gradient approaches zero at the sub-central point 
(see Appendix B and Fig. B.1).  This establishes \CFP\ as a fairly 
commonly-distributed species in diffuse and translucent molecular gas.  


In Fig. 2 (see Table 2) we broadly compared the velocity distribution of \CFP\ 
with those of several hydrocarbons and \hcop\ tracing the total column density, 
with HNC that appears strongly only when the number and column densities 
are higher locally, and with HCO, an interface species that is known not 
to penetrate dense gas.  The hydrocarbons and \hcop\ have higher 
integrated opacity on the high-velocity side of the profiles where column 
density aggregates as the result of  velocity-crowding in portions of the
line of sight at the smallest galactocentric radii.  HNC behaves oppositely, 
peaking in the 40 \kms\ feature  which is the strongest individual cloud
feature  seen along the line of sight.  In this regard, \CFP\ most closely 
resembles HNC. 

In Fig. 3 we compared the column densities of \CFP, \hcop, \cch\ and HNC 
more quantitatively along all the sightlines where \CFP\ has been observed in 
absorption.  The 33 \kms\ cloud toward W49N has molecular column densities 
most nearly similar to the sightline toward \bll\ at \AV\ = 1 mag, but 
only 40\% as much \CFP.  The 39 \kms\ feature toward W49N is similar to 
the higher-column density translucent sightline toward 3C111 (\AV\ $\ga 4$ mag), 
including \CFP.  N(\CFP) shows a quadratic dependence on N(\cch) and
most closely resembles HNC in its detailed behaviour.

As shown in Table 4 the relative abundance  X(\CFP) = N(\CFP)/N(\HH) 
observed in absorption toward W49N are $ 3.3\times10^{-10}$ for the 
gas around 39 \kms\ and $1.4\pm0.16\times10^{-10}$ for the other four 
velocity intervals in which \CFP\ was detected.  To interpret these
results we embedded the fluorine chemistry in a model of the heating, 
cooling and \HH\ and CO formation in diffuse and translucent gas using 
a recent remeasurement of the F $+$ \HH $\rightarrow$ HF + H formation 
reaction.  Using CH as a surrogate for \HH\ with X(CH) = $3.5\times10^{-8}$
the models correctly reproduce the observed column density and relative
abundance of HF with X(HF) = N(HF)/N(\HH) $= 1.1 \times 10^{-8}$.  However, 
they also predict N(\CFP)/N(\HH) $= 5.3 \times 10^{-10}$ and 
N(\CFP)/N(HF) $\approx 1/20$ which is too large by factors of four for 
\CFP\ seen over most of the velocity range seen toward W49, or by 
40\% for the feature around 39 \kms.

Given the correctness of the predictions for HF, we hypothesized that the 
discrepancy in the predicted \CFP\ abundance and \CFP/HF ratio in the models
results from  the absence of a significant \CFP\ destruction process.
This process is presumably suppressed when the \CFP\ relative abundance 
is unusually high.  We surmised that photodissociation of \CFP\ is important 
and we showed, by examination of the recombination rates in the models, that 
a photodissociation rate $\Gamma \ga 10^{-9}$ \ps (comparable
to that of CH) could cause a substantial decrease in the \CFP\ abundance,
introducing number and column density dependences that otherwise are largely
absent.  However, the free space photodissocation rate has recently been
calculated and found to be  $\Gamma = 2\times 10^{-10}$ \ps, far too
small to explain the observed discrepancies.  In any case, uncertainties
remain in the adopted \CFP\ formation and recombination rates, which
may need revision.

In Fig. 4 we compared the optical depth profiles of C\p\ and HF, the 
species that form  \CFP, with that of \CFP\ itself.  The profile of 
$\tau$(C\p) varies little across the peaks in $\tau$(HF) and $\tau$(\CFP), 
owing to blending of atomic and molecular gas and somewhat poorer 
velocity resolution, with, perhaps, a decreased C\p\
fraction in the high-column density cloud.  If C\p\ is the dominant form
of gas-phase carbon it dominates both the formation and destruction of 
\CFP, leaving \CFP\ to trace N(HF) and N(\HH).  In this experiment
along the complex and heavily-blended low-latitude line of sight to
W49N, \CFP\ signaled the coexistence of C\p\ and HF, but primarily 
traced the peaks of the \HH\ distribution.  

\begin{table*}
\caption[]{Optical depth integrals over selected velocity intervals$^a$}
{
\small
\begin{tabular}{cccccccc}
\hline
Vel. range&C$_3$\HH&\hcop&HNC&{red \CFP}&HCO&\hthcop&CH\\
\kms&\kms&\kms&\kms&\kms&\kms&\kms&\kms \\
\hline
23.9..31.3&0.520(0.040)&---&0.000(0.017)&0.0000(0.0027)&0.0000(0.0121)&---&0.035(0.019)\\
31.6..35.7&1.380(0.040)&1.950(0.018)&0.270(0.018)&0.0110(0.0024)&0.0540(0.0096)&0.0000(0.0112)&1.144(0.014)\\
35.7..42.0&3.140(0.070)&6.210(0.074)&1.970(0.036)&0.0514(0.0029)&0.1322(0.0115)&0.1240(0.0140)&2.040(0.019)\\
42.1..50.0&0.660(0.040)&0.840(0.016)&0.000(0.023)&0.0000(0.0032)&0.0325(0.0126)&0.0000(0.0154)&0.861(0.017)\\
50.0..56.8&1.570(0.045)&3.980(0.025)&0.400(0.022)&0.0161(0.0030)&0.0867(0.0118)&0.0000(0.0144)&1.660(0.017)\\
56.8..61.5&2.780(0.054)&7.270(0.070)&1.200(0.023)&0.0255(0.0026)&0.2410(0.0102)&0.1234(0.0123)&2.150(0.018)\\
61.5..66.1&2.370(0.053)&4.490(0.036)&0.880(0.021)&0.0218(0.0025)&0.1410(0.0097)&0.0649(0.0118)&1.830(0.017)\\
66.1..70.3&0.628(0.330)&0.810(0.013)&0.110(0.017)&0.0000(0.0024)&0.0275(0.0095)&0.0000(0.0113)&0.670(0.013)\\
\hline
\end{tabular}}
\\
$^a$ Quantities in parenthesis are $1\sigma$ statistical error estimates, see Sect. 2
\end{table*}


\begin{table*}
\caption[]{Line frequencies and opacity-column density conversion factors}
{
\small
\begin{tabular}{ccccccccccc}
\hline
& C$_3$\HH&\hcop&HNC&\CFP&HCO&\hthcop& \cch& HF & C$^+$ & CH\\
\hline
frequency$^a$ & 85.339 &89.189 & 90.664 & 102.588 & 86.671  &86.754& 87.317 & 1232.476 
& 1900.537  & 532.774 \\
factor$^b$ &4.36E12 & 1.12E12 & 1.76E12 & 1.35E13 & 2.26E13 & 1.13E12
 & 6.52E13& 2.404E12  &1.40E17 &3.64E13\\
\hline
\end{tabular}}
\\
$^a$ GHz \\
$^b$ units for all entries are $\pcc$ (\kms)$^{-1}$ \\
\end{table*}

\begin{table*}
\caption[]{Relative Abundances$^a$}
{
\small
\begin{tabular}{cccccccc}
\hline
 Vel. range&C$_3$\HH&\hcop&HNC&\CFP&HCO&C$_2$H&HF\\
\kms&&&&&&&\\
\hline
31.6..35.7&1.48E-09&1.84E-09&3.99E-10&1.25E-10&5.08E-11&3.99E-08&1.30E-08\\
35.7..42.1&1.66E-09&3.28E-09&1.63E-09&3.26E-10&6.98E-11&4.56E-08&1.29E-08\\
42.0..50.0&1.03E-09&1.05E-09& $<$1.36E-10& $<$1.44E-10 &4.07E-11&2.89E-08&6.77E-09\\
50.0..56.8&1.23E-09&2.58E-09&4.08E-10&1.26E-10&5.62E-11&2.97E-08&9.57E-09\\
56.8..61.5&1.45E-09&3.64E-09&9.45E-10&1.54E-10&1.21E-10&3.60E-08&1.20E-08\\
61.5..66.1&1.43E-09&2.64E-09&8.14E-10&1.54E-10&8.30E-11&3.17E-08&9.68E-09\\
66.1..70.3&1.25E-09&1.30E-09&2.78E-10& $<$1.39E-10&4.42E-11&2.66E-08&8.59E-09\\
\hline
Sum$^b$&1.41E-09&2.66E-09&7.89E-10&1.57E-10&7.43E-11&3.54E-08&1.08E-08\\
Mean$^c$&1.36E-09(0.21)&2.33E-09(0.98)&7.46E-10(5.05)&1.77E-10(0.85)
&6.65E-11(2.82) &3.41E-08(0.68)&1.04E-08(0.24)\\
\hline
\end{tabular}}
\\
$^a$ Quantities tabulated are N(X)/N(\HH) where N(\HH) = N(CH)/$3.5\,10^{-8}$ \\
$^b$ Entries labelled Sum are ratios of the total column density over all intervals to the total CH \\
$^c$ Entries labelled Mean are the average(rms) of the detections in the body of the table.
\end{table*}

\begin{appendix}

\section{Extraction of the \CFP\ line profile}

As noted in Sect. 2.1, We extracted the \CFP\ profile in the presence of an
overlying H66$\epsilon$ recombination line.  We have confidence in the
identification of this extraneous feature by comparison with maps we made of
the more isolated H53$\gamma$ line, whose 38 \kms\ linewidth is
the same as that of the gaussian profile fitted to extract the \CFP\
profile.

Shown in Fig A.1 is the observed 
continuum-subtracted profile, as derived from a linear 
fit to the spectral baseline well outside the velocity interval shown.  
A gaussian profile with FWHM 38.2$\pm1.6$ \kms was fit to the spectrum 
using spectral channels 
within velocity intervals chosen to coincide with regions of weak HF 
absorption, in order to avoid removing any broad underlying component
of \CFP\ absorption.  The fitted gaussian is shown in red and the velocity 
intervals used for fitting are outlined as blue dashed rectangles.  The 
sum of the gaussian and linear fits was then used to derive the final 
continuum-subtracted profile from the original data.  The continuum flux 
at the observed position was 3.47 Jy and the rms line/continuum noise 
at zero-absorption in the final continuum-subtracted spectrum was 
$2.4\,10^{-3}$.   

We tested the reliability of the fit by adding to the data varying
amounts of absorption mimicking the optical depth spectrum of HF
\citep{GerLev+12}, to which \CFP\ is closely related, and then 
observing whether the added absorption was present in the derived 
\CFP\ profile after gaussian decomposition.  This was the case as
long as the velocity intervals used in the profile decomposition were
those in which HF absorption toward W49N is weak.  We are confident
that no broad component of \CFP\ absorption was removed by the 
profile decomposition, although such a feature may be well be present
below the levels of detectability of the experiment.

\begin{figure}
\includegraphics[height=6.6cm]{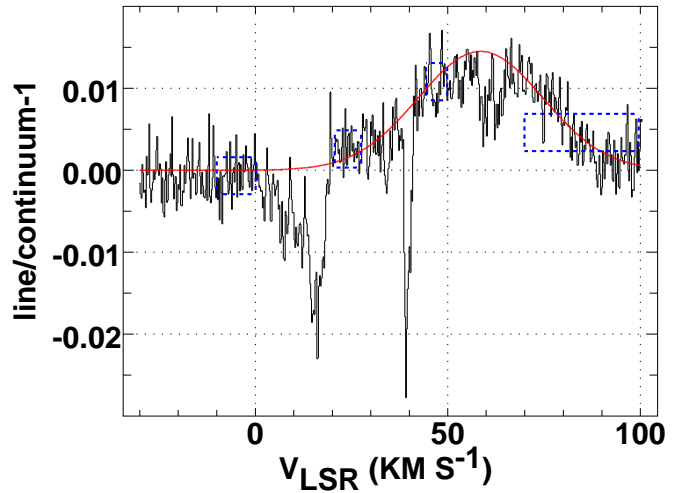}
  \caption[]{\CFP\ and recombination line profiles observed toward W49N.
  The solid black histogram is the continuum-subtracted profile using a 
 linear spectral baseline fit to  signal-free channels well outside the region 
 shown.   The red line shows the gaussian profile fitted to the 
 recombination line emission over the spectrum segments outlined in blue, 
 which were chosen to coincide with regions of small HF, hence \CFP\,  
 absorption.}
\end{figure}

\section{Line of sight kinematics}

\begin{figure}
\includegraphics[height=8.7cm]{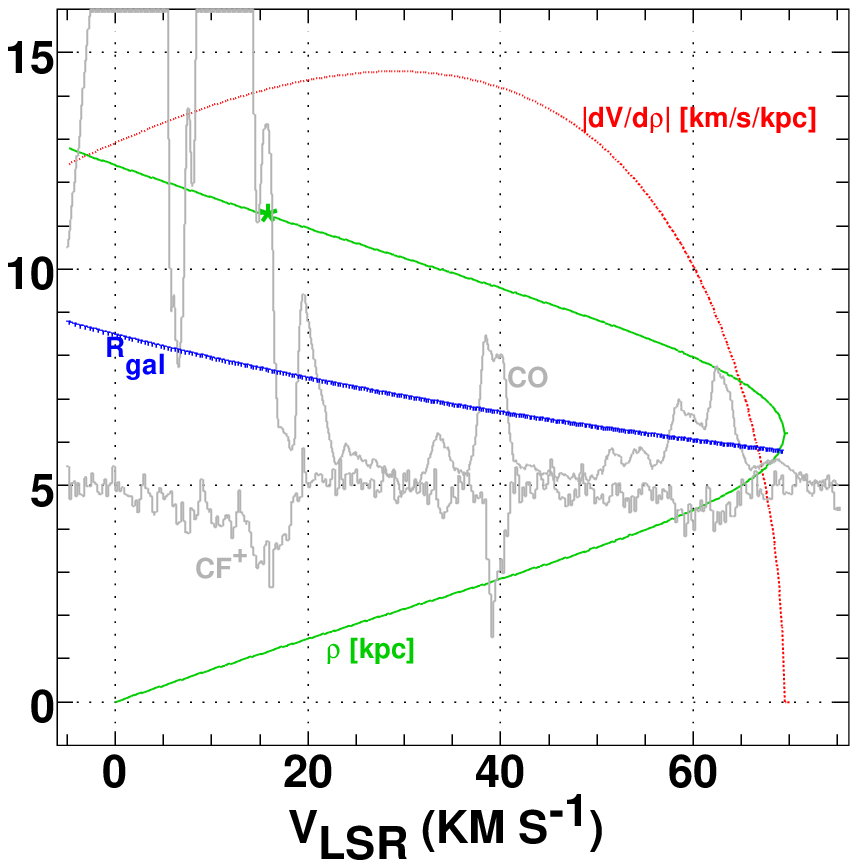}
  \caption[]{W49N line of sight kinematics in a nutshell.  The mottled blue line shows
the galactocentric radius R at each velocity assuming a flat rotation curve with 
$\Theta(R) = \Theta(R_0) = 220 $\kms, R$_0$ = 8.5 kpc.  The solid green curve 
shows the line of sight distance or distances $\rho$ at each velocity and 
the distance to W49 $\rho$, 11.1$\pm0.8$ kpc \citep{ZhaRei13} is illustrated 
by an an asterisk. The dotted red curve shows the line of sight velocity 
gradient dV/d$\rho$ due to Galactic rotation.  Shown superposed to illustrate the 
observed range of spectral 
features are the CO profile in units of Kelvins, offset but not scaled, and the \CFP\
absorption spectrum scaled upward a factor of 100 and offset, both from Fig. 1.
}
\end{figure}

The line of sight kinematics toward W49 are illustrated in very compact form in Fig. B.1
for a flat rotation curve, as noted in the figure caption. W49N is at the far kinematic
distance only slightly inside the Solar Circle so the same path is sampled in both
emission and absorption.  The maximum velocities of features 
observed in emission and absorption correspond well to that expected from the model
velocity field.  The so-called terminal velocity feature at $v \ga 60$ \kms\ occurs
as material over a wide range of line of sight distance collects in a narrowed
velocity range owing to the smaller line of sight velocity gradient.  The velocity
range up to +20 \kms\ includes the nearest 1.5 \kpc\ along the line of sight and
constrains the absorption to arise inward of 7.5 kpc in galactocentric distance..

\end{appendix}

\begin{acknowledgements}
The National Radio Astronomy Observatory is a facility of the National Science Foundation 
operated under cooperative agreement by Associated Universities, Inc.
IRAM is operated by CNRS (France), MPG (Germany) and IGN (Spain). 
This work was supported by the CNRS program ``Physique et Chimie du Milieu
Interstellaire'' (PCMI). PG acknowledges funding by the European Research 
Council (Starting Grant 3DICE, grant agreement 336474, PI: V. Wakelam).
We thank Fabrice Dayou and Evelyne Roueff for communicating the result 
of their calculation of the \CFP\ photdissociation rate prior to 
publication.

\end{acknowledgements}

\bibliographystyle{apj}


\begin{thebibliography}{35}
\expandafter\ifx\csname natexlab\endcsname\relax\def\natexlab#1{#1}\fi

\bibitem[{{Burgh} {et~al.}(2010){Burgh}, {France}, \& {Jenkins}}]{BurFra+10}
{Burgh}, E.~B., {France}, K., \& {Jenkins}, E.~B. 2010, Astroph. J., 708, 334

\bibitem[{{Burton}(1971)}]{Bur71}
{Burton}, W.~B. 1971, A\&A, 10, 76

\bibitem[{{Fechtenbaum} {et~al.}(2015){Fechtenbaum}, {Bontemps}, {Schneider},
  {Csengeri}, {Duarte-Cabral}, {Herpin}, \& {Lefloch}}]{FecBon+15}
{Fechtenbaum}, S., {Bontemps}, S., {Schneider}, N., {Csengeri}, T.,
  {Duarte-Cabral}, A., {Herpin}, F., \& {Lefloch}, B. 2015, A\&A, 574, L4

\bibitem[{{Gerin} {et~al.}(2010){Gerin}, {de Luca}, {Goicoechea}, {Herbst},
  {Falgarone}, {Godard}, {Bell}, {Coutens}, {Ka{\'z}mierczak}, {Sonnentrucker},
  {Black}, {Neufeld}, {Phillips}, {Pearson}, {Rimmer}, {Hassel}, {Lis},
  {Vastel}, {Boulanger}, {Cernicharo}, {Dartois}, {Encrenaz}, {Giesen},
  {Goldsmith}, {Gupta}, {Gry}, {Hennebelle}, {Hily-Blant}, {Joblin},
  {Ko{\l}os}, {Kre{\l}owski}, {Mart{\'{\i}}n-Pintado}, {Monje}, {Mookerjea},
  {Perault}, {Persson}, {Plume}, {Salez}, {Schmidt}, {Stutzki}, {Teyssier},
  {Yu}, {Contursi}, {Menten}, {Geballe}, {Schlemmer}, {Morris}, {Hatch},
  {Imram}, {Ward}, {Caux}, {G{\"u}sten}, {Klein}, {Roelfsema}, {Dieleman},
  {Schieder}, {Honingh}, \& {Zmuidzinas}}]{GerdeL+10}
{Gerin}, M., {de Luca}, M., {Goicoechea}, J.~R., {Herbst}, E., {Falgarone}, E.,
  {Godard}, B., {Bell}, T.~A., {Coutens}, A., {Ka{\'z}mierczak}, M.,
  {Sonnentrucker}, P., {Black}, J.~H., {Neufeld}, D.~A., {Phillips}, T.~G.,
  {Pearson}, J., {Rimmer}, P.~B., {Hassel}, G., {Lis}, D.~C., {Vastel}, C.,
  {Boulanger}, F., {Cernicharo}, J., {Dartois}, E., {Encrenaz}, P., {Giesen},
  T., {Goldsmith}, P.~F., {Gupta}, H., {Gry}, C., {Hennebelle}, P.,
  {Hily-Blant}, P., {Joblin}, C., {Ko{\l}os}, R., {Kre{\l}owski}, J.,
  {Mart{\'{\i}}n-Pintado}, J., {Monje}, R., {Mookerjea}, B., {Perault}, M.,
  {Persson}, C., {Plume}, R., {Salez}, M., {Schmidt}, M., {Stutzki}, J.,
  {Teyssier}, D., {Yu}, S., {Contursi}, A., {Menten}, K., {Geballe}, T.~R.,
  {Schlemmer}, S., {Morris}, P., {Hatch}, W.~A., {Imram}, M., {Ward}, J.~S.,
  {Caux}, E., {G{\"u}sten}, R., {Klein}, T., {Roelfsema}, P., {Dieleman}, P.,
  {Schieder}, R., {Honingh}, N., \& {Zmuidzinas}, J. 2010, A\&A, 521, L16

\bibitem[{{Gerin} {et~al.}(2011){Gerin}, {Ka{\'z}mierczak}, {Jastrzebska},
  {Falgarone}, {Hily-Blant}, {Godard}, \& {de Luca}}]{GerKaz+11}
{Gerin}, M., {Ka{\'z}mierczak}, M., {Jastrzebska}, M., {Falgarone}, E.,
  {Hily-Blant}, P., {Godard}, B., \& {de Luca}, M. 2011, A\&A, 525, A116

\bibitem[{{Gerin} {et~al.}(2012){Gerin}, {Levrier}, {Falgarone}, {Godard},
  {Hennebelle}, {Le Petit}, {De Luca}, {Neufeld}, {Sonnentrucker}, {Goldsmith},
  {Flagey}, {Lis}, {Persson}, {Black}, {Goicoechea}, \& {Menten}}]{GerLev+12}
{Gerin}, M., {Levrier}, F., {Falgarone}, E., {Godard}, B., {Hennebelle}, P.,
  {Le Petit}, F., {De Luca}, M., {Neufeld}, D., {Sonnentrucker}, P.,
  {Goldsmith}, P., {Flagey}, N., {Lis}, D.~C., {Persson}, C.~M., {Black},
  J.~H., {Goicoechea}, J.~R., \& {Menten}, K.~M. 2012, Royal Society of London
  Philosophical Transactions Series A, 370, 5174

\bibitem[{{Gerin} {et~al.}(2015){Gerin}, {Ruaud}, {Goicoechea}, {Gusdorf},
  {Godard}, {de Luca}, {Falgarone}, {Goldsmith}, {Lis}, {Menten}, {Neufeld},
  {Phillips}, \& {Liszt}}]{GerRua+15}
{Gerin}, M., {Ruaud}, M., {Goicoechea}, J.~R., {Gusdorf}, A., {Godard}, B., {de
  Luca}, M., {Falgarone}, E., {Goldsmith}, P., {Lis}, D.~C., {Menten}, K.~M.,
  {Neufeld}, D., {Phillips}, T.~G., \& {Liszt}, H. 2015, A\&A, 573, A30

\bibitem[{{Godard} {et~al.}(2010){Godard}, {Falgarone}, {Gerin}, {Hily-Blant},
  \& {de Luca}}]{GodFal+10}
{Godard}, B., {Falgarone}, E., {Gerin}, M., {Hily-Blant}, P., \& {de Luca}, M.
  2010, A\&A, 520, A20

\bibitem[{{Guzm{\'a}n} {et~al.}(2012{\natexlab{a}}){Guzm{\'a}n}, {Pety},
  {Gratier}, {Goicoechea}, {Gerin}, {Roueff}, \& {Teyssier}}]{GuzPet+12}
{Guzm{\'a}n}, V., {Pety}, J., {Gratier}, P., {Goicoechea}, J.~R., {Gerin}, M.,
  {Roueff}, E., \& {Teyssier}, D. 2012{\natexlab{a}}, A\&A, 543, L1

\bibitem[{{Guzm{\'a}n} {et~al.}(2012{\natexlab{b}}){Guzm{\'a}n}, {Roueff},
  {Gauss}, {Pety}, {Gratier}, {Goicoechea}, {Gerin}, \& {Teyssier}}]{GuzRou+12}
{Guzm{\'a}n}, V., {Roueff}, E., {Gauss}, J., {Pety}, J., {Gratier}, P.,
  {Goicoechea}, J.~R., {Gerin}, M., \& {Teyssier}, D. 2012{\natexlab{b}}, A\&A,
  548, A94

\bibitem[{{Helfer} \& {Blitz}(1997)}]{HelBli97}
{Helfer}, T.~T. \& {Blitz}, L. 1997, Astroph. J., 478, 233

\bibitem[{{Indriolo} {et~al.}(2013){Indriolo}, {Neufeld}, {Seifahrt}, \&
  {Richter}}]{IndNeu+13}
{Indriolo}, N., {Neufeld}, D.~A., {Seifahrt}, A., \& {Richter}, M.~J. 2013,
  Astroph. J., 764, 188

\bibitem[{{Liszt} \& {Lucas}(2001)}]{LisLuc01}
{Liszt}, H. \& {Lucas}, R. 2001, A\&A, 370, 576

\bibitem[{{Liszt}(1995)}]{Lis95}
{Liszt}, H.~S. 1995, Astroph. J., 442, 163

\bibitem[{{Liszt}(2007)}]{Lis07CO}
---. 2007, A\&A, 476, 291

\bibitem[{{Liszt}(2015)}]{Lis15HD}
---. 2015, Astroph. J., 799, 66

\bibitem[{{Liszt} {et~al.}(1993){Liszt}, {Braun}, \& {Greisen}}]{LisBra+93}
{Liszt}, H.~S., {Braun}, R., \& {Greisen}, E.~W. 1993, Astron. J., 106, 2349

\bibitem[{{Liszt} {et~al.}(2014){Liszt}, {Pety}, {Gerin}, \&
  {Lucas}}]{LisPet+14}
{Liszt}, H.~S., {Pety}, J., {Gerin}, M., \& {Lucas}, R. 2014, A\&A, 564, A64

\bibitem[{{Liszt} {et~al.}(2010){Liszt}, {Pety}, \& {Lucas}}]{LisPet+10}
{Liszt}, H.~S., {Pety}, J., \& {Lucas}, R. 2010, A\&A, 518, A45+

\bibitem[{{Lucas} \& {Liszt}(1996)}]{LucLis96}
{Lucas}, R. \& {Liszt}, H.~S. 1996, A\&A, 307, 237

\bibitem[{{Lucas} \& {Liszt}(2000)}]{LucLis00C2H}
---. 2000, A\&A, 358, 1069

\bibitem[{{McElroy} {et~al.}(2013){McElroy}, {Walsh}, {Markwick}, {Cordiner},
  {Smith}, \& {Millar}}]{McEWal+13}
{McElroy}, D., {Walsh}, C., {Markwick}, A.~J., {Cordiner}, M.~A., {Smith}, K.,
  \& {Millar}, T.~J. 2013, A\&A, 550, A36

\bibitem[{{Neufeld} {et~al.}(2006){Neufeld}, {Schilke}, {Menten}, {Wolfire},
  {Black}, {Schuller}, {M{\"u}ller}, {Thorwirth}, {G{\"u}sten}, \&
  {Philipp}}]{NeuSch+06}
{Neufeld}, D.~A., {Schilke}, P., {Menten}, K.~M., {Wolfire}, M.~G., {Black},
  J.~H., {Schuller}, F., {M{\"u}ller}, H.~S.~P., {Thorwirth}, S., {G{\"u}sten},
  R., \& {Philipp}, S. 2006, Astrophysics, 454, L37

\bibitem[{{Neufeld} \& {Wolfire}(2009)}]{NeuWol09}
{Neufeld}, D.~A. \& {Wolfire}, M.~G. 2009, \apj, 706, 1594

\bibitem[{{Pety}(2005)}]{Pet05}
{Pety}, J. 2005, in SF2A-2005: Semaine de l'Astrophysique Francaise, ed.
  F.~{Casoli}, T.~{Contini}, J.~M. {Hameury}, \& L.~{Pagani}, 721

\bibitem[{{Rachford} {et~al.}(2002){Rachford}, {Snow}, {Tumlinson}, {Shull},
  {Blair}, {Ferlet}, {Friedman}, {Gry}, {Jenkins}, {Morton}, {Savage},
  {Sonnentrucker}, {Vidal-Madjar}, {Welty}, \& {York}}]{RacSno+02}
{Rachford}, B.~L., {Snow}, T.~P., {Tumlinson}, J., {Shull}, J.~M., {Blair},
  W.~P., {Ferlet}, R., {Friedman}, S.~D., {Gry}, C., {Jenkins}, E.~B.,
  {Morton}, D.~C., {Savage}, B.~D., {Sonnentrucker}, P., {Vidal-Madjar}, A.,
  {Welty}, D.~E., \& {York}, D.~G. 2002, Astroph. J., 577, 221

\bibitem[{{Sheffer} {et~al.}(2008){Sheffer}, {Rogers}, {Federman}, {Abel},
  {Gredel}, {Lambert}, \& {Shaw}}]{SheRog+08}
{Sheffer}, Y., {Rogers}, M., {Federman}, S.~R., {Abel}, N.~P., {Gredel}, R.,
  {Lambert}, D.~L., \& {Shaw}, G. 2008, Astroph. J., 687, 1075

\bibitem[{{Snow} {et~al.}(2007){Snow}, {Destree}, \& {Jensen}}]{SnoDes+07}
{Snow}, T.~P., {Destree}, J.~D., \& {Jensen}, A.~G. 2007, Astroph. J., 655, 285

\bibitem[{{Sofia} {et~al.}(2004){Sofia}, {Lauroesch}, {Meyer}, \&
  {Cartledge}}]{SofLau+04}
{Sofia}, U.~J., {Lauroesch}, J.~T., {Meyer}, D.~M., \& {Cartledge}, S.~I.~B.
  2004, Astroph. J., 605, 272

\bibitem[{{Sonnentrucker} {et~al.}(2010){Sonnentrucker}, {Neufeld}, {Phillips},
  {Gerin}, {Lis}, {de Luca}, {Goicoechea}, {Black}, {Bell}, {Boulanger},
  {Cernicharo}, {Coutens}, {Dartois}, {Ka{\'z}mierczak}, {Encrenaz},
  {Falgarone}, {Geballe}, {Giesen}, {Godard}, {Goldsmith}, {Gry}, {Gupta},
  {Hennebelle}, {Herbst}, {Hily-Blant}, {Joblin}, {Ko{\l}os}, {Kre{\l}owski},
  {Mart{\'{\i}}n-Pintado}, {Menten}, {Monje}, {Mookerjea}, {Pearson},
  {Perault}, {Persson}, {Plume}, {Salez}, {Schlemmer}, {Schmidt}, {Stutzki},
  {Teyssier}, {Vastel}, {Yu}, {Caux}, {G{\"u}sten}, {Hatch}, {Klein}, {Mehdi},
  {Morris}, \& {Ward}}]{SonNeu+10}
{Sonnentrucker}, P., {Neufeld}, D.~A., {Phillips}, T.~G., {Gerin}, M., {Lis},
  D.~C., {de Luca}, M., {Goicoechea}, J.~R., {Black}, J.~H., {Bell}, T.~A.,
  {Boulanger}, F., {Cernicharo}, J., {Coutens}, A., {Dartois}, E.,
  {Ka{\'z}mierczak}, M., {Encrenaz}, P., {Falgarone}, E., {Geballe}, T.~R.,
  {Giesen}, T., {Godard}, B., {Goldsmith}, P.~F., {Gry}, C., {Gupta}, H.,
  {Hennebelle}, P., {Herbst}, E., {Hily-Blant}, P., {Joblin}, C., {Ko{\l}os},
  R., {Kre{\l}owski}, J., {Mart{\'{\i}}n-Pintado}, J., {Menten}, K.~M.,
  {Monje}, R., {Mookerjea}, B., {Pearson}, J., {Perault}, M., {Persson}, C.~M.,
  {Plume}, R., {Salez}, M., {Schlemmer}, S., {Schmidt}, M., {Stutzki}, J.,
  {Teyssier}, D., {Vastel}, C., {Yu}, S., {Caux}, E., {G{\"u}sten}, R.,
  {Hatch}, W.~A., {Klein}, T., {Mehdi}, I., {Morris}, P., \& {Ward}, J.~S.
  2010, A\&A, 521, L12

\bibitem[{{Sonnentrucker} {et~al.}(2015){Sonnentrucker}, {Wolfire}, {Neufeld},
  {Flagey}, {Gerin}, {Goldsmith}, {Lis}, \& {Monje}}]{SonWol+15}
{Sonnentrucker}, P., {Wolfire}, M., {Neufeld}, D.~A., {Flagey}, F., {Gerin},
  M., {Goldsmith}, P., {Lis}, D., \& {Monje}, R. 2015, Astroph. J., 799,
  submitted

\bibitem[{{Tizniti} {et~al.}(2014){Tizniti}, {Le Picard}, {Lique},
  {Berteloite}, {Canosa}, {Alexander}, \& {Sims}}]{TizLeP+14}
{Tizniti}, M., {Le Picard}, S.~D., {Lique}, F., {Berteloite}, C., {Canosa}, A.,
  {Alexander}, M.~H., \& {Sims}, I.~R. 2014, Nat. Chem., 6, 141

\bibitem[{{Van Dishoeck} \& {Black}(1986)}]{VanBla86}
{Van Dishoeck}, E.~F. \& {Black}, J.~H. 1986, Astrophys. J., Suppl. Ser., 62,
  109

\bibitem[{{Weselak} {et~al.}(2010){Weselak}, {Galazutdinov}, {Beletsky}, \&
  {Kre{\l}owski}}]{WesGal+10}
{Weselak}, T., {Galazutdinov}, G.~A., {Beletsky}, Y., \& {Kre{\l}owski}, J.
  2010, Mon. Not. R. Astron. Soc., 402, 1991

\bibitem[{{Zhang} {et~al.}(2013){Zhang}, {Reid}, {Menten}, {Zheng},
  {Brunthaler}, {Dame}, \& {Xu}}]{ZhaRei13}
{Zhang}, B., {Reid}, M.~J., {Menten}, K.~M., {Zheng}, X.~W., {Brunthaler}, A.,
  {Dame}, T.~M., \& {Xu}, Y. 2013, Astroph. J., 775, 79

\end{thebibliography}

\end{document}